\documentclass[useAMS,usenatbib]{mn2e}
\usepackage{epsfig}
\usepackage{times}
\usepackage[fleqn]{amsmath}
\usepackage{amssymb}
\usepackage{amsbsy}

\title[Power spectrum extraction for the LOFAR EoR experiment]{Power
spectrum extraction for redshifted 21-cm epoch of reionization
experiments: the LOFAR case}

\author[G. J. A. Harker et al.]{Geraint Harker,$^{1,2}$\thanks{E-mail:
  geraint.harker@colorado.edu} Saleem Zaroubi,$^{3}$ Gianni
  Bernardi,$^{4}$ Michiel A. Brentjens,$^{3}$ \newauthor A. G. de
  Bruyn,$^{3,5}$ Benedetta Ciardi,$^{6}$ Vibor Jeli\'c,$^{3}$ Leon
  V. E. Koopmans,$^{3}$ \newauthor Panagiotis Labropoulos,$^{3}$
  Garrelt Mellema,$^{7}$ Andr\'e Offringa,$^{3}$ V. N. Pandey,$^{3}$
  \newauthor Andreas H. Pawlik,$^{8,9}$ Joop Schaye,$^{8}$ Rajat
  M. Thomas$^{10}$ and Sarod Yatawatta$^{3}$\\ $^{1}$Center for
  Astrophysics and Space Astronomy, 389 UCB, University of Colorado,
  Boulder, CO 80309-0389, USA\\ $^{2}$NASA Lunar Science Institute,
  NASA Ames Research Center, Moffett Field, CA, USA\\ $^{3}$Kapteyn
  Astronomical Institute, University of Groningen, PO Box 800, 9700AV
  Groningen, the Netherlands\\ $^{4}$Harvard--Smithsonian Center for
  Astrophysics, 60 Garden Street, Cambridge, MA 02138, USA\\
  $^{5}$ASTRON, Postbus 2, 7990AA Dwingeloo, the Netherlands\\
  $^{6}$Max-Planck Institute for Astrophysics,
  Karl-Schwarzschild-Stra\ss e 1, 85748 Garching, Germany\\
  $^{7}$Department of Astronomy and Oskar Klein Centre for
  Cosmoparticle Physics, AlbaNova, Stockholm University, SE-106 91
  Stockholm, Sweden\\ $^{8}$Leiden Observatory, Leiden University, PO
  Box 9513, 2300RA Leiden, the Netherlands\\ $^{9}$Department of
  Astronomy, University of Texas, Austin, TX 78712, USA\\
  $^{10}$Institute for the Mathematics and Physics of the Universe
  (IPMU), The University of Tokyo, Chiba 277-8582, Japan}

\begin{document}

\date{\today}

\maketitle

\begin{abstract}
One of the aims of the Low Frequency Array (LOFAR) Epoch of
Reionization (EoR) project is to measure the power spectrum of
variations in the intensity of redshifted 21-cm radiation from the EoR.
The sensitivity with which this power spectrum can be estimated depends
on the level of thermal noise and sample variance, and also on the
systematic errors arising from the extraction process, in particular
from the subtraction of foreground contamination. We model the
extraction process using realistic simulations of the cosmological
signal, the foregrounds and noise, and so estimate the sensitivity of
the LOFAR EoR experiment to the redshifted 21-cm power spectrum.
Detection of emission from the EoR should be possible within 360 hours
of observation with a single station beam. Integrating for longer, and
synthesizing multiple station beams within the primary (tile) beam,
then enables us to extract progressively more accurate estimates of the
power at a greater range of scales and redshifts. We discuss different
observational strategies which compromise between depth of observation,
sky coverage and frequency coverage. A plan in which lower frequencies
receive a larger fraction of the time appears to be promising. We also
study the nature of the bias which foreground fitting errors induce on
the inferred power spectrum, and discuss how to reduce and correct for
this bias.  The angular and line-of-sight power spectra have different
merits in this respect, and we suggest considering them separately in
the analysis of LOFAR data.
\end{abstract}


\section{Introduction}\label{sec:intro}

Studying 21-cm radiation from hydrogen at high redshifts
\citep*{FIE58,FIE59,HOG79,SCO90,KUM95,MAD97} promises to be interesting
for several reasons.  Fluctuations in intensity are sourced partly by
density fluctuations, measurements of which may allow rather tight
constraints on cosmological parameters \citep{MAO08}.  They are also
sourced by variations in the temperature and ionized fraction of the
gas, which means that 21-cm studies may provide information on early
sources of ionization and heating, such as stars or mini-QSOs. The
period during which the gas undergoes the transition from being largely
neutral to largely ionized is known as the Epoch of Reionization
\citep[EoR; e.g.][]{LOE01,BEN06,FOB06}, while the period beforehand is
sometimes known as the cosmic dark ages.  While the latter has perhaps
the best potential to give clean constraints on cosmology, the
instruments becoming available in the near future are not expected to
be sensitive enough at the appropriate frequencies to study this epoch
interferometrically. Several, though, are hoped to be able to study the
EoR (e.g.\ GMRT,\footnotemark\ MWA,\footnotemark\ LOFAR,\footnotemark\
21CMA,\footnotemark\ PAPER,\footnotemark\ SKA\footnotemark), but even
so, their sensitivity is not expected to be sufficient to make high
signal-to-noise images of the 21-cm emission in the very near future.
We seek, instead, a statistical detection of a cosmological 21-cm
signal, with the most widely studied statistic being the power spectrum
(e.g. \citealt{MOR04}; \citealt{BAR05}; \citealt{MCQ06};
\nocite{BMH06,BOW07}Bowman, Morales \& Hewitt 2006, 2007;
\citealt{PRI07}; \citealt{BAR09}; \citealt{LID08}; \citealt{PRI08};
\citealt{SET08}). Our aim in this paper is to test how well the 21-cm
power spectrum can be extracted from data collected with the Low
Frequency Array (LOFAR), which is currently under construction. While
this is a general-purpose observatory, the EoR project, being one of
LOFAR's Key Science Projects, has helped to drive the design of the
instrument. We give some details on parameters of the instrument which
are relevant to EoR observations in Section~\ref{subsec:instresp}.

The quality of extraction is affected by several factors: the
observational strategy and the length of observations, which affect the
volume being studied and the level of thermal noise; the array design
and layout; the foregrounds from Galactic and extragalactic sources,
and the methods used to remove their influence from the data
(presumably by exploiting their assumed smoothness as a function of
frequency; see e.g. \citealt{SHA99}; \citealt{DIM02}; \citealt{OH03};
\citealt*{ZAL04}); excision of radio-frequency interference (RFI) and
radio recombination lines; and, for example, the quality of
polarization and total intensity calibration for instrumental and
ionospheric effects. We will not study RFI or calibration here. We
will, however, use simulations of the cosmological signal (CS), the
foregrounds, the instrumental response and the noise to generate
synthetic data cubes -- i.e.\ the intensity of 21-cm emission as a
function of position on the sky and observing frequency -- and then
attempt to extract the 21-cm power spectrum from these cubes. We
generate data cubes realistic enough so that we can test different
observing strategies and methods of subtracting the foregrounds, and
look at the effect on the inferred power spectrum.
\footnotetext[1]{Giant Metrewave Telescope,
http://www.gmrt.ncra.tifr.res.in/}\footnotetext[2]{Murchison Widefield
Array, http://www.haystack.mit.edu/ast/arrays/mwa/}\footnotetext[3]{Low
Frequency Array, http://www.lofar.org/}\footnotetext[4]{21 Centimeter
Array, http://web.phys.cmu.edu/\~{}past/}\footnotetext[5]{Precision
Array to Probe the EoR,
http://astro.berkeley.edu/\~{}dbacker/eor/}\footnotetext[6]{Square
Kilometre Array, http://www.skatelescope.org/}

We devote the following section to describing the construction of the
data cubes and giving a brief description of their constituent parts.
Then, in Section~\ref{sec:ext} we discuss the extraction of the 21-cm
power spectrum from the cubes, including our method for subtracting
the foregrounds. In Section~\ref{sec:sens} we present our estimates of
the sensitivity of LOFAR to the 21-cm power spectrum, and discuss the
character of the statistical and systematic errors on these estimates.
We conclude in Section~\ref{sec:conc} by offering some thoughts on
what these results suggest about the merits of different observing
strategies and extraction techniques.

\section{Simulations}\label{sec:sims}

\subsection{Cosmological signal and foregrounds}\label{subsec:csandfg}

We test the quality and sensitivity of our power spectrum extraction
using synthetic LOFAR data cubes, which have various components.  The
first is the redshifted 21-cm signal which is simulated as described by
\citet{THO09}.  The starting point for this is a dark matter simulation
of $512^3$ particles in a cube with sides of comoving length $200\
h^{-1}\ \mathrm{Mpc}$. The sides thus have twice the length of the
simulations exhibited by \citet{THO09} and used in our previous work on
LOFAR EoR signal extraction \citep{SKE_09,NONPAR_09}, allowing us to
probe larger scales. The assumed cosmological parameters are
\hbox{($\Omega_\mathrm{m}$, $\Omega_\Lambda$, $\Omega_\mathrm{b}$, $h$,
$\sigma_8$, $n$)}$=$\hbox{(0.238, 0.762, 0.0418, 0.73, 0.74, 0.951)},
where all the symbols have their usual meaning. This leads to a minimum
resolved halo mass of around $3\times 10^{10}\ h^{-1}\
\mathrm{M}_\odot$. Dark matter haloes are populated with sources whose
properties depend on some assumed model. For this paper we assume the
`quasar-type' source model of \citet{THO09}, which is better suited to
this simulation than one assuming stellar sources owing to the
relatively low resolution, which raises the minimum resolved halo mass.
The topology and morphology of reionization is different compared to a
simulation with a stellar source model, and the power spectrum is also
slightly different. We might expect quasar reionization to allow an
easier detection than stellar reionization, since the regions where the
sources are found are larger and more highly clustered, producing
larger fluctuations in the signal. This paper is concerned with the
extraction of the power in general, however, and the precise source
properties are not expected to affect our conclusions since the fitting
appears to be relatively unaffected by the difference in the source
model \citep{NONPAR_09}.

Given the source properties, the pattern of ionization is computed
using a one-dimensional radiative transfer code \citep{THO08}, which
allows realizations to be generated very rapidly in a large volume. If
the spin temperature is sufficiently large, as we assume here, the
differential brightness temperature between 21-cm emission and the CMB
is given by \citep{MAD97,CIA03}
\begin{equation}
\frac{\delta
  T_\mathrm{b}}{\mathrm{mK}}=39h(1+\delta)x_\mathrm{HI}\left(\frac{\Omega_\mathrm{b}}{0.042}\right)\left[\left(\frac{0.24}{\Omega_\mathrm{m}}\right)\left(\frac{1+z}{10}\right)\right]^\frac{1}{2}\
  ,
\label{eqn:dtb}
\end{equation}
where $\delta$ is the matter density contrast, $x_\mathrm{HI}$ is the
neutral hydrogen fraction, and the current value of the Hubble
parameter, \hbox{$H_0=100h\ \mathrm{km}\ \mathrm{s}^{-1}\
\mathrm{Mpc}^{-1}$}. The series of periodic simulation snapshots from
different times is converted to a continuous observational cube
(position on the sky versus redshift or observational frequency) using
the scheme described by \citet{THO09}.  In brief, the emission in each
snapshot is calculated in redshift space (i.e.\ taking into account
velocities along the line of sight, which cause redshift-space
distortions). Then, at each observing frequency at which an output is
required, the signal is calculated by interpolating between the
appropriate simulation boxes.  We use frequencies between $121.5$ and
$200\ \mathrm{MHz}$, so we have a `frequency cube' of size $200\
h^{-1}\ \mathrm{Mpc}\ \times\ 200\ h^{-1}\ \mathrm{Mpc}\ \times\ 78.5\
\mathrm{MHz}$. To approximate the field of view of a LOFAR station,
however, we use a square observing window of $5\degr\times 5\degr$,
which corresponds to comoving distances of around $600\ h^{-1}\
\mathrm{Mpc}$ at the redshifts corresponding to EoR observations. We
therefore tile copies of the frequency cube in the plane of the sky to
fill this observing window, and interpolate the resulting data cube
onto a grid with $256\times 256\times 158$ points. This simplified
treatment of the field of view implicitly assumes that the station
beam is equal to unity everywhere within a square window of
frequency-independent angular size, and zero outside. Since we plan to
use only the top part of the primary beam for EoR measurements, the
sensitivity will vary relatively slowly across the field of view. Our
simulations of the CS restrict us to examining angular modes much
smaller than the size of the beam in any case, and so the main effect
of this simplification is to slightly decrease the overall level of
noise compared to a more accurate beam model. As we progress to using
larger simulations of the CS, which let us examine more angular modes,
the effects of the primary beam will become more important and will be
included in future work.

\begin{figure}
  \begin{center}
    \leavevmode \psfig{file=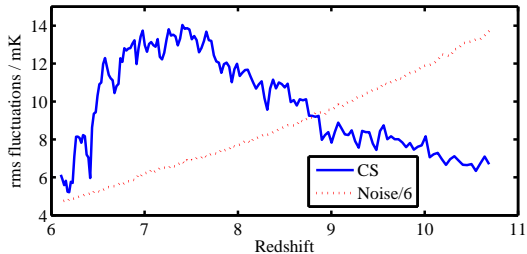,width=8cm} \caption{The rms
    fluctuation in differential brightness temperature, calculated at
    the resolution of LOFAR, in our simulation of the cosmological
    signal (CS) is shown as a function of redshift (solid line). For
    comparison, we show the rms noise for an observing time of 600
    hours per frequency channel, scaled down by a factor of 6 (dotted
    line). Note that the vertical axis scale does not start at
    zero.}\label{fig:rms}
  \end{center}
\end{figure}

The rms variation in differential brightness temperature in each slice
of this data cube is shown as a function of redshift in
Fig.~\ref{fig:rms}.
This rms is calculated at the resolution of LOFAR, which will be
around $4\ \mathrm{arcmin}$ for EoR observations, depending on
frequency. Note that the rms fluctuation does not drop to zero by the
lowest redshift in this simulation, indicating that reionization is
not complete there. This delay in reionization comes about because the
source properties are the same as for our earlier, higher-resolution
simulations, which contain more resolved haloes (i.e.\ the minimum
resolved halo mass is lower). The larger simulations therefore have
fewer sources per unit volume. Such late reionization appears
unrealistic given current observational constraints \citep*[e.g.\
][and references therein]{FAN06}, and means that extracting the power
spectrum at low redshift may be more difficult in reality than we
would predict using these simulations.  The most stringent test of our
power spectrum extraction occurs at higher redshift, however, since
this corresponds to lower observing frequencies at which the noise
(shown in Fig.~\ref{fig:rms}) and the foregrounds are larger. The
power spectrum evolves less strongly at high redshift, and so we
expect this simulation to perform reasonably well there compared to
high resolution simulations. It may even be slightly conservative,
since H\textsc{ii} regions at high redshift may increase the strength
of fluctuations at some scales.

We use the foreground simulations of \citet{JEL08}. These incorporate
contributions from Galactic diffuse synchrotron and free-free emission,
and supernova remnants. They also include unresolved extragalactic
foregrounds from radio galaxies and radio clusters. We assume, however,
that point sources bright enough to be distinguished from the
background, either within the field of view or outside it, have been
removed perfectly from the data. Observations of foregrounds at $150\
\mathrm{MHz}$ at low latitude \citep{BER09} indicate that these
simulations fairly describe the properties of the diffuse foregrounds.

\subsection{Instrumental response}\label{subsec:instresp}

LOFAR is a radio interferometer which is planned to have fields of
antennas (stations) in several European countries. Its core, however,
is near the village of Exloo in the Netherlands, and it is the stations
in the core area (and perhaps some nearby `remote stations') which will
be used for EoR observations. Each station contains two types of
antenna: low-band antennas (LBA), optimized for $30$--$80\
\mathrm{MHz}$, and high-band antennas (HBA) optimized for $120$--$240\
\mathrm{MHz}$. The LBAs will not be sensitive enough for redshifted
21-cm work, so we will be concerned only with the HBAs. EoR
observations are expected to take place below approximately $190\
\mathrm{MHz}$ (above $z=6.48$).

To improve the {\it uv} coverage (at the expense of increasing the
workload of the supercomputer which acts as LOFAR's correlator), within
each LOFAR core station the HBA antennas are distributed into two
semi-stations, each of which is then treated is an independent station.
The antennas are collected into tiles, each of which is a grid of
$4\times 4$ dual dipoles.  A semi-station consists of 24 such tiles,
arranged in a filled circle.  A remote station has all 48 of its HBA
tiles collected into a single circle. Each pair of stations provides us
with one baseline.

\begin{figure}
  \begin{center}
    \leavevmode \psfig{file=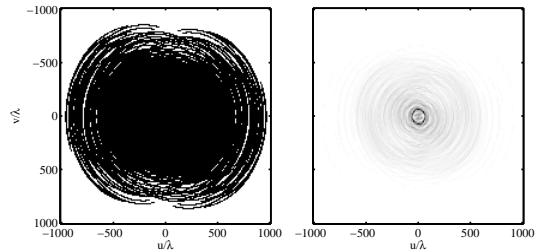,width=8cm} \caption{Assumed {\it uv} coverage at $150\ \mathrm{MHz}$ (left panel): black cells are those containing at least one observation, i.e. those having $S(u,v)>0$. The right panel shows the density of points in the {\it uv} plane, on a linear greyscale.}\label{fig:sfunuv}
  \end{center}
\end{figure}
To include the effects of the instrumental response of LOFAR we define
a sampling function $S(u,v)$ which describes how densely the
interferometer baselines sample Fourier space over the course of an
observation, such that $1/\sqrt{S}$ is proportional to the noise on the
measurement of the Fourier transform of the sky in each {\it uv} cell.
In general this sampling function is frequency-dependent, but we
examine the effect of this dependence by comparing to a situation in
which we assume the {\it uv} coverage is the same at all frequencies.
This situation could be approximated in practice by not using data at
{\it uv} points for which there is no coverage at some frequencies.
This would involve discarding approximately 20 per cent of the data
(from the outer part of the {\it uv} plane at high frequencies, and
from the inner part at low frequencies), increasing the level of noise
and reducing the resolution at high frequencies. Throughout this paper,
$S(u,v)$ is computed under the assumption that 24 dual stations in the
core and the first ring of LOFAR are used to observe a window at a
declination of $90\degr$. We assume noise levels appropriate to an
observation at the zenith, however. The final LOFAR layout is likely to
include fewer dual stations, and EoR observations will use some of the
more central remote stations, but we will not investigate different
configurations in this paper. The sampling function and {\it uv}
coverage at $150\ \mathrm{MHz}$, at which the frequency-dependent and
frequency-independent sampling functions match, are shown in
Fig.~\ref{fig:sfunuv}. The {\it uv} tracks are for a four-hour
observation. We summarize some of the parameters of our simulated
observations using this array layout in Table~\ref{table:obspars}.
\begin{table}
\caption{Parameters of our synthetic observations and assumed array
layout.} \label{table:obspars}
\begin{center}
\begin{tabular}{ll}
\hline
Total effective area at $150\ \mathrm{MHz}$ & $2.46\times 10^4\ \mathrm{m}^2$ \\
Image noise for a 300 hour observation & $78\ \mathrm{mK}$ \\
\ with $1\ \mathrm{MHz}$ bandwidth at $150\ \mathrm{MHz}$ & \\
Frequency coverage & $121.5$--$200\ \mathrm{MHz}$ \\
Frequency channel width & $0.5\ \mathrm{MHz}$ \\
Station beam field of view & $5\degr\times 5\degr$ \\
Number of instantaneous baselines & $48\times 47$ \\
Spatial resolution at $150\ \mathrm{MHz}$ & $\approx 4\ \mathrm{arcmin}$ \\
\hline
\end{tabular}
\end{center}
\end{table}

To simulate our data in the {\it uv} plane we perform a
two-dimensional Fourier transform on the image of the foregrounds and
signal at each frequency, and multiply by a mask (the {\it uv}
coverage) which is unity at grid points in Fourier space ({\it uv}
cells) where $S(u,v)>0$, and is zero elsewhere. At this point we add
uncorrelated complex Gaussian noise with an rms proportional to
$1/\sqrt{S}$ to the cells within the mask. We can then return to the
image plane by performing an inverse two-dimensional Fourier transform
at each frequency. This two-dimensional Fourier relationship between
the {\it uv} and image plane only holds approximately for long
integrations with a LOFAR-type array, but we use it here since it
allows considerable simplification. The overall normalization of the
level of noise at each frequency is chosen to match the expected rms
noise of single-channel images. Part of the aim of this paper is to
check the effect of different levels of noise on power spectrum
extraction.  For reference, we assume that 300 hours of observation of
one EoR window with one synthesized beam with LOFAR will give noise
with an rms of $78\ \mathrm{mK}$ on an image using $1\ \mathrm{MHz}$
bandwidth at $150\ \mathrm{MHz}$. Although this is a somewhat
conservative choice, it offsets the assumption of a uniform primary
beam within the field of view we are considering, since a more
realistic model for the primary beam would produce a noise rms that
increased towards the edge of the field of view. The level of noise
varies with frequency, being related to the system temperature which
we assume to be $T_\mathrm{sys}=140+60(\nu/300\ \mathrm{MHz})^{-2.55}\
\mathrm{K}$.

A much more detailed account of the calculation of noise levels and
the effects of instrumental corruption for the LOFAR EoR project may
be found in \citet{LAB09}.

\section{Extraction}\label{sec:ext}

\subsection{The problem of extraction}\label{subsec:problem}

In this paper, the main limitation on the quality of power spectrum
extraction which we will consider is the subtraction of astrophysical
foregrounds. One difficulty encountered in this subtraction is simply
that the fluctuations in the foregrounds are much larger than those in
the CS: a subtraction algorithm must ensure that features due to the
signal are not mistaken for relatively tiny features in the
foregrounds.  A second difficulty is the presence of noise, which
limits the accuracy and precision with which we are able to measure
the foregrounds, and hence the accuracy with which we can subtract
them.  The relative importance of these two effects changes with
scale, since the power spectra of the foregrounds, signal and noise do
not have the same shape.

Our foreground subtraction relies on the foregrounds being spectrally
smooth, i.e.\ lacking small-scale features in the frequency direction.
Any small-scale features are put down to noise or signal. Large-scale
features due to the CS are more difficult to recover, since they can
easily be confused with foreground features.  The difficulty of
recovering the large-scale power is exacerbated because the
fluctuations in the foregrounds become larger compared to the noise and
the signal, making the problem of overfitting more severe.

At small scales, the noise is more of an issue: its power spectrum
becomes much larger compared to the foregrounds and signal, making the
latter impossible to pick out.  The scale-dependence of the
contaminants means that there is a `sweet spot': a range of scales at
which both the foregrounds and the noise are small enough relative to
the CS for the prospects for signal extraction to be good.

This fact has implications for choosing an observational strategy for
the LOFAR EoR experiment, because we must trade off the depth of
observation against sky and frequency coverage.  A deep observation of
a small area allows foreground fits of higher quality, and is
especially beneficial for the recovery of small-scale power.  It
limits the size and number of modes which we can sample, however,
which is especially damaging for the errors on the recovered
large-scale power.  Conversely, increasing the size of the area
surveyed beats down sample variance and may allow us to probe larger
scales, though note that in the case of radio interferometry the
length of the shortest baselines sets an upper limit on the size of
the available modes. This increase in area is only useful, however, if
the noise levels are low enough to allow foreground fitting to take
place.

Examining this trade-off is one of the aims of this work.  Before
doing so, we first outline the procedures we have used to fit the
foregrounds.

\subsection{Fitting procedure}\label{subsec:fitproc}

As we mentioned in Section~\ref{sec:sims}, we consider both the case
in which the {\it uv} coverage of the observations depends on
observing frequency, and the idealized case in which it does not. For
the latter, we always fit the foregrounds in the image-space frequency
cube using the Wp smoothing method \citep{MAC93,MAC95} described in
detail in \citet{NONPAR_09} and summarized in
Section~\ref{subsubsec:wp}. This method requires the specification of
a parameter, $\lambda$, which governs the level of regularization:
larger values impose a smoother solution. We use $\lambda=0.5$ for our
image-space fitting, since we found this to work well for extracting
the rms \citep{NONPAR_09}. Before fitting, we reduce the resolution of
the images, combining blocks of $4\times 4$ pixels together to
generate a $64\times 64\times 158$ data cube. Since the unbinned
pixels are smaller than a resolution element of LOFAR (the binned
pixels are slightly larger), and since the relative contribution of
the noise increases at small scales, this does not discard spatial
scales at which we can usefully extract information, but does increase
the quality of the fit, reducing bias.

When the {\it uv} coverage is frequency-dependent, however, fitting in
image space becomes problematic, since spatial fluctuations are
converted to fluctuations in the frequency direction, as illustrated
by, for example, \citet*{BOW09} and \citet{LIU09b}. Instead, we leave
the data cube in Fourier space [or, to be more precise,
$(u,v,\nu)$-space, since we do not transform along the frequency
direction], and fit the foregrounds as a function of frequency at each
{\it uv} point before subtracting them and generating images. The real
and imaginary parts are fit separately, using inverse-variance weights
to take account of the fact that the noise properties change as a
function of frequency.  This implies that if a point in the {\it uv}
plane is not sampled at a particular frequency, then it has zero weight
and does not contribute to the fit. This is therefore similar to the
method proposed by \citet{LIU09b}. We discard `lines of sight' in
Fourier space in which the weight is non-zero for fewer than ten
points, since the foregrounds are not well constrained here and we
would merely introduce noise into the residual images.

This leaves the problem of which method to use to perform the fitting
in Fourier space. Choosing a method is more awkward than in image
space, since the mean contribution from foregrounds, noise and signal
varies across the {\it uv} plane. It may be optimal to vary the
parameters of a fitting method according to the position in the {\it
uv} plane. None the less, we obtain reasonable results simply using a
third-order polynomial in frequency to fit the real and imaginary
parts at each point in the plane. We have also used Wp smoothing to
fit the foregrounds in the {\it uv} plane. This gives us the freedom
to vary the smoothing parameter, $\lambda$, across the plane. Near the
origin (i.e.\ corresponding to large spatial scales) little
regularization is required, since the contribution from the
foregrounds is much larger than that from the signal or the noise and
so they are well measured. Toward the edges of the plane we need to
make stronger assumptions about the smoothness of the foregrounds to
avoid overfitting, and so we make the value of $\lambda$ larger.
Finding a `natural choice' for $\lambda$ is somewhat awkward (see
\citealt{NONPAR_09} for further discussion), so at present we choose a
mean value of $\lambda$ which gives reasonable results, and vary it
between lines of sight by making it inversely proportional to the
mean, $\bar{c}$, of the fitting weights of points along that line of
sight. Specifically, we use $\lambda(u,v)=280/\bar{c}(u,v)$, where
$c(u,v,\nu_i)=\sqrt{S(u,v,\nu_i)}/\sigma^\mathrm{im}(\nu_i)$ and
$\sigma^\mathrm{im}(\nu_i)$ is the rms image noise at frequency
$\nu_i$ expressed in kelvin. Since the noise is typically a few tenths
of a kelvin, and $S$ has values ranging up to around $2.5\times 10^5$,
we end up with $\lambda\approx 15$ at the edge of the {\it uv} plane
and $\lambda\approx 0.03$ near the centre, for an integration of 300
hours. The results are not sensitive to the precise normalization of
$\lambda$.

\subsubsection{Wp smoothing}\label{subsubsec:wp}

Wp smoothing is a non-parametric fitting method which appears to be
very suitable for fitting the spectrally smooth foregrounds in EoR
data sets. It was developed for general cases by \citet{MAC93,MAC95},
and we have described an algorithm for using it for fitting EoR
foregrounds in a previous paper \citep{NONPAR_09}. We will briefly
outline its principles here.

The aim is to fit a function $f(x)$ to a series of points
$\{(x_i,y_i)\}$ subject to a constraint on the number of inflection
points in the function, and on the integrated change of curvature away
from the inflection points.  More precisely, define the function
$h_f(x)$ by
\begin{equation}
f''(x)=s_f(x-w_1)(x-w_2)\ldots (x-w_{n_w})\mathrm{e}^{h_f(x)}
,\label{hfdef}
\end{equation}
where $s_f=\pm 1$ and $w_1,\ldots ,w_{n_w}$ are the inflection points.
The function $f$ we wish to find is that which minimizes
\begin{equation}
\sum_{i=1}^n \rho_i(y_i-f(x_i)) +
\lambda\int_{x_1}^{x_n}h_f'(t)^2\mathrm{d}t\ ,
\label{eqn:minimizes}
\end{equation}
where the function $\rho_i$, which takes as its argument the
difference $\delta=y_i-f(x_i)$ between the fitting function and the
data points, penalizes the fitting function if it strays too far from
the data.  We opt to use a least-squares fit, with
$\rho_i(\delta)=c_i/(2 \delta^2)$ where $c_i$ is a weight. Our choice
for $c_i$ is given above. The parameter $\lambda$ controls the
relative importance of the least-squares term and the regularization
term, with larger values giving heavier smoothing.

\citet{MAC93,MAC95} derives an ordinary differential equation and
appropriate boundary conditions such that the solution is the function
$f$ which we require. We solve it by discretizing it to give an
algebraic system which we solve using standard methods. It is possible
to perform a further minimization over the number and position of the
inflection points, but we have found that solutions with no inflection
points fit the EoR foregrounds well, so we do not require this extra
step.

\subsection{Power spectrum estimation}\label{subsec:psest}

Once we have fit the foregrounds, we subtract the fit to leave a
residual data cube which has as its components the cosmological signal,
the noise and any fitting errors.  We will mainly be concerned with the
spherically averaged three-dimensional power spectra of the residuals
and their components. These are calculated within some sub-volume of
the full data cube (for example, a slice $8\ \mathrm{MHz}$ thick) by
computing the power in cells and then averaging it in spherical annuli
to give band-power estimates. Each cell contributes only to the
annulus in which its centre lies, i.e.\ we ignore the fact that the
cells have non-zero size. The annuli are logarithmically spaced, but
because we plot the power against the mean value of $k$ for cell
centres lying within an annulus, the points in figures may not be
exactly logarithmically spaced. Rather than showing the raw power, in
our figures we plot the quantity
$\Delta^2(k)=\mathcal{V}k^3P(k)/(2\pi^2)$ \citep[or the analogous one-
or two-dimensional quantity: see e.g.][]{KAI91}, where $\mathcal{V}$ is
the volume. This is usually called the dimensionless power spectrum
when dealing with the spectrum of overdensities, though in this case it
has the dimensions of temperature squared. $\Delta^2(k)$ is then the
contribution to the temperature fluctuations from modes in a
logarithmic bin around the wavenumber $k$.

Different systematic effects are important for modes along and across
the line of sight, however. For this reason we also calculate the
two-dimensional power spectrum perpendicular to the line of sight
(i.e.\ the angular power spectrum, but expressed as a function of
cosmological wavenumber $k$) and the one-dimensional power spectrum
along the line of sight. We estimate the two-dimensional power spectrum
at a particular frequency by averaging the power in annuli. Estimates
calculated from one frequency band tend to be rather noisy, so we
usually average the power spectrum across several frequency bands to
give a less noisy estimate. In the one-dimensional case we simply
calculate the one-dimensional power spectrum for each line of sight
with no additional binning (producing points linearly spaced in
$k$), then average these spectra across all $64^2$ lines of sight
[$256^2$ lines of sight in the case of the cubes fit in $(u,v,\nu)$
space] to give an estimate for the whole volume. Typically we consider
a volume only $\sim 8\ \mathrm{MHz}$ deep, so that the CS does not
evolve too much within the volume.

To see more clearly the contribution to the power spectrum of the
residuals from its different components, we write the residuals in
Fourier space as
\begin{equation}
r(\boldsymbol{k})=s(\boldsymbol{k})+n(\boldsymbol{k})+\epsilon(\boldsymbol{k})\ ,
\label{eqn:rdef}
\end{equation}
where $s$ is the cosmological signal, $n$ is the noise and $\epsilon$
is the fitting error. Then the power spectrum is given by
\begin{align}
P^r(k)={}&\langle
r(\boldsymbol{k})r(\boldsymbol{k})^\ast\rangle_{|\boldsymbol{k}|=k}\label{eqn:prdef}\\
={}&P^s(k)+P^n(k)+P^\epsilon(k)\nonumber\\
&+\langle\epsilon(\boldsymbol{k})[s(\boldsymbol{k})+n(\boldsymbol{k})]^\ast+[s(\boldsymbol{k})+n(\boldsymbol{k})]\epsilon(\boldsymbol{k})^\ast\rangle_{|\boldsymbol{k}|=k}\label{eqn:prexp}
\end{align}
where the subscript indicates that the averaging takes place over a
shell in $k$-space, and the superscripts label the power spectra of the
different components. The equality on the second line follows because
the signal and noise are uncorrelated so their cross-terms average to
zero. We cannot assume, however, that the fitting errors are
uncorrelated with the signal or noise, which gives rise to the final
term in angle brackets, which may be either positive or negative. We
may usually expect it to be negative, since we fit away some of the
signal and noise, reducing the size of the residuals. If it is large
enough, the power spectrum of the residuals may even fall below the
power spectrum of the input CS, especially at scales where the noise
power is small.

If we ignore the fitting errors, we may estimate the power spectrum of
the CS by computing the power spectrum of the residuals, then
subtracting the expected power spectrum of the noise. In this case, we
can make a relatively straightforward estimate of the error on the
extracted power spectrum, as we see in
Section~\ref{subsubsec:staterrors}. We have assumed here that the
expected power spectrum of the noise is known to reasonable accuracy.
In fact, we will not be able compute it accurately enough {\it a
priori} for real LOFAR data: it must instead be estimated through
observation. It should be possible to do so by differencing adjacent,
narrow frequency channels (much narrower than those in the simulations
used here, where the data have been binned into $0.5\ \mathrm{MHz}$
channels: the estimate would have to be carried out before this level
of binning, using channels of perhaps $10\ \mathrm{kHz}$). Studying
this in more detail in the context of the LOFAR EoR experiment must be
the subject of future work, though note that this approach has already
been applied to characterize the noise in low frequency foreground
observations made with the Westerbork telescope \citep{BER10}, the
GMRT \citep*{ALI08} and PAPER \citep{PAR09}.

\subsubsection{Statistical errors}\label{subsubsec:staterrors}

The statistical errors on the extracted power spectrum include
contributions from the noise and from sample variance.  Considering
first the noise, in the $i^\mathrm{th}$ Fourier cell the real and
imaginary parts of the contribution to the gridded visibility from the
noise, $V^n_i$, are Gaussian-distributed, with mean zero and variance
$\sigma_i^2$ (say), which is known. Then $|V^n_i|^2$ is exponentially
distributed with mean $2\sigma_i^2$ and variance $4\sigma_i^4$. We may
estimate the power spectrum at some wavenumber $k$ by computing
\begin{equation}
\langle P^n(k)\rangle=\frac{1}{m_k}\sum_{i=1}^{m_k}|V^n_i|^2
\end{equation}
where the sum is over all cells within an annulus near $k$. If the
number of cells in the annulus is sufficiently large, the error on
this estimate is approximately Gaussian-distributed, and we estimate
it as $\langle P^n(k)\rangle /\sqrt{m_k}$, assuming that the different
cells are independent and using the fact that the variance of
$|V^n_i|^2$ is the square of its mean. This error translates into an
error on the final extracted power spectrum, and can be reduced either
by integrating longer on the same patch of sky (to reduce
$\sigma_i^2\sim 1/\tau$ where $\tau$ is the observing time) or by
spending the time observing a wider area to increase the number of
accessible modes, increasing $m_k$. In the latter case, the error only
decreases as $1/\sqrt{\tau}$.

Though this estimate of the error is useful as a guide for how the
errors behave as the observational parameters change, a more accurate
error bar can be computed in a Monte Carlo fashion by looking at the
dispersion between independent realizations of the noise, and this is
how we compute the errors in practice. Although the analytic estimate
is reasonable, it tends to underestimate the errors at large scales
and overestimate them at small scales.

The power spectrum of our simulation of the CS is calculated similarly
to the power spectrum of the noise. In this case, the error $\langle
P^\mathrm{CS}(k)\rangle /\sqrt{m_k}$ represents the error on our final
estimate of the power spectrum due to sample variance, and can only be
reduced by sampling more modes (increasing $m_k$). Unlike the noise,
the fluctuations in the CS are not Gaussian, and so an analytic
estimate of the error is likely to be less accurate. This should not
matter too much at small scales where in any case the error on our
extracted power spectrum is dominated by noise, but on larger scales
the sample variance becomes important. At present we do not have
enough different realizations of the CS to simulate the errors more
realistically: as noted in Section~\ref{sec:sims} we must already tile
copies of a single simulation to fill a LOFAR field of view, which
limits the range of scales we can realistically study. These estimates
should therefore be considered an illustration of how we expect the
errors to change as we vary our observational strategy, rather than a
definitive calculation, which is reasonable given the other
simplifications we have made (e.g.\ adopting a square field of view
rather than a realistic primary beam shape). Error bars on our
extracted power spectra are computed by adding the noise and sample
variance errors in quadrature.

\subsubsection{Systematic errors}\label{subsubsec:syserr}

The terms involving fitting errors on the right-hand side of
equation~\eqref{eqn:prexp} will bias our estimate of the power
spectrum of the CS unless they can be accurately corrected for, and so
contribute to a systematic error. When analysing LOFAR data it may be
possible to estimate the size of these terms using simulations similar
to the ones used in this paper. \citet{BOW09} have estimated them for
simulations of MWA data through a `subtraction characterization
factor' $f_s(k)=\langle P^s(k)\rangle/P^s(k)$. By fitting cubes which
include different realizations of the CS and noise, it should also be
possible to reflect the statistical error introduced by making such a
correction in the error bars. In this paper we do not make this
correction, however: it would be accurate by construction and hence
quite uninformative. Instead we plot $\langle
P^s(k)\rangle=P^r(k)-\langle P^n(k)\rangle $ to illustrate the level
of bias we may expect to see if no correction is made. Our error bars
will then reflect errors due only to the sample variance and the
noise. If the estimated power falls below the true power, we use the
estimate of sample variance from the true power, since this gives a
more realistic view of what the estimate of the sample variance would
be if we made a correction for the fitting bias.

We expect any estimate of the bias, or of the statistical error
introduced by correcting for the bias, to be rather uncertain, since
it may depend strongly on the shape of the foregrounds, which is
unknown to the required level of accuracy {\it a priori}, and on the
details of the fitting procedure used. It is none the less
straightforward to estimate them for a specific foreground model and
fitting procedure.

\subsubsection{Cross-correlation}\label{subsubsec:crosscorr}

As an alternative to calculating a residual power spectrum and
then subtracting a thermal noise power spectrum, we could obtain the
extracted power spectrum through cross-correlation.  That is, we could
split an observing period into two sub-epochs, subtract the foregrounds
from each and then cross-correlate the two. Following the approach
taken to derive Equation~\eqref{eqn:prexp}, we can write the residual
in each of the two epochs as
\begin{equation}
r_i(\boldsymbol{k})=s(\boldsymbol{k})+n_i(\boldsymbol{k})+\epsilon_i(\boldsymbol{k})\ ,
\label{eqn:rdefi}
\end{equation}
where the signal $s(k)$ is the same for the two cases and $i\in
\{1,2\}$ labels the epoch. Then
\begin{equation}
\langle r_1r_2^\ast\rangle = P^s + \langle s\epsilon_2^\ast\rangle + \langle \epsilon_1s^\ast\rangle + \langle \epsilon_1\epsilon_2^\ast\rangle\ , \label{eqn:ccexp}
\end{equation}
where the $\boldsymbol{k}$-dependence is implicit, the angle brackets
again indicate an average over a shell in $k$-space, and cross-terms
involving the noise vanish. If the fitting errors are sufficiently
small, this cross-correlation immediately provides us with an estimate
of the desired power spectrum.

This estimator has some apparent advantages. Firstly, we do not have to
know the thermal noise power spectrum to calculate it (though an
estimate of the thermal noise is required to compute error bars).
Secondly, we do not expect it to yield negative estimates of the power,
as may happen when using Equation~\eqref{eqn:prexp}. More generally, at
scales where the noise is larger than the signal or the fitting errors,
we would expect the bias of this estimator to be much smaller than for
the one involving autocorrelations, since the cross-terms involving $n$
and $\epsilon$ on the right-hand side of Equation~\eqref{eqn:prexp} do
not appear.

It is not without disadvantages, however.  If we split the observation
into two epochs, the lower signal-to-noise in each epoch will degrade
the foreground fitting, increasing the size of the $\epsilon$ terms.
If, instead, the foreground fitting is done on the full dataset before
dividing it into different epochs, then the cross-terms involving $n$
and $\epsilon$ can no longer be assumed to vanish.

We have conducted preliminary tests of the cross-correlation method and
found that it gives comparable results to the autocorrelation method at
scales where the fitting bias is small enough for either estimate to be
useful. We reiterate, however, that it is assumed here that the thermal
noise power spectrum is known accurately, which unfairly favours using
the autocorrelation.  We defer further comparison of the two methods
until we have looked further into how well the noise power spectrum can
be estimated from observations. In this paper, all our extracted power
spectra are computed by subtracting the noise power spectrum from the
residual power spectrum. We would not expect our broad conclusions to
change if we were to use cross-correlation instead.

\section{Sensitivity estimates}\label{sec:sens}

\subsection{Comparison of fitting methods}\label{subsec:diffmeth}

\begin{figure}
  \begin{center}
    \leavevmode \psfig{file=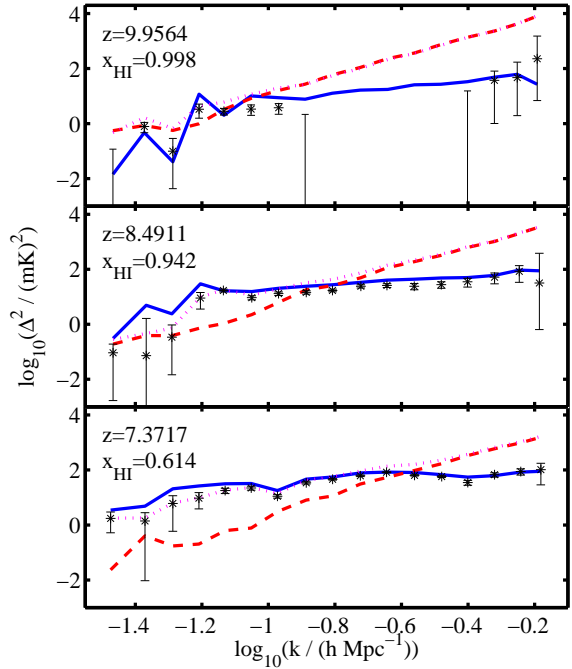,width=8cm}
    \caption{Power spectra of the input CS (solid line), the noise
    (dashed line), the residuals (dotted line) and the extracted
    signal (points with error bars) at three different redshifts. Here
    we assume the {\it uv} coverage is frequency-independent, so the
    foreground fitting is done using Wp smoothing in the image
    plane. The noise level is consistent with 300 hours of observation
    per frequency bin on a single window, using one station beam. The
    redshift shown in each panel is the central redshift of an $8\
    \mathrm{MHz}$ slice from the frequency cube. This frequency
    interval corresponds to $\Delta z=0.63$, $0.48$ and $0.37$ for the
    top, middle and bottom panel respectively. From top to bottom, the
    mean neutral fraction in each slice, $\bar{x}_\mathrm{HI}$, is
    0.9976, 0.9416 and 0.6140. The missing points in the top panel
    correspond to $k$ bins at which the power spectrum of the
    residuals falls below the power spectrum of the noise, so that we
    would infer an unphysical, negative signal
    power.}\label{fig:threez300bin4}
  \end{center}
\end{figure}

Examples of extracted power spectra at three different redshifts, for
slices $8\ \mathrm{MHz}$ thick, are given in
Fig.~\ref{fig:threez300bin4} (points with error bars).
From top to bottom, the central redshift of the slice used in each
panel is 9.96, 8.49 and 7.37, while the mean neutral fraction
$\bar{x}_\mathrm{HI}$ in each slice is 0.998, 0.942 and 0.614,
respectively.

For comparison, we also show the power spectrum of the noiseless CS
cube (solid line), the noise (dashed line) and the residuals after
fitting (dotted line). The extracted power spectrum is the difference
between the residual and noise power spectra, and would be equal to
the noiseless CS power spectrum if there were no foregrounds. For this
figure we use a frequency-independent {\it uv} coverage, so the
foreground fitting is carried out in the low-resolution image cube. A
noise level consistent with 300 hours of observation per frequency bin
of a single ($5\degr \times 5\degr$) window using a single station
beam is assumed. It may not be possible to observe the entire
frequency range simultaneously, and it may have to be split into two
or three segments (e.g. of $32\ \mathrm{MHz}$ width) only one of which
can be observed at once. If we have to use two such segments, then the
300 hours of observation per frequency bin translates to 600 hours of
total observing time. This is a somewhat pessimistic scenario for the
quality of data we may collect after one year of EoR observations with
LOFAR, since it is hoped that several station beams can be correlated
simultaneously to cover the top of the primary (tile) beam, allowing a
larger field of view to be mapped out more quickly. It may also be
possible to trade off the number of beams against the width of the
frequency window, or to spend different amounts of time on different
parts of the frequency range. None the less, the assumptions of
Fig.~\ref{fig:threez300bin4} provide a useful baseline against which
we can compare results for deeper observations or for more realistic
(frequency-dependent) {\it uv} coverage. It also illustrates the main
features we see in many of our extracted spectra.

For the lowest-redshift slice (bottom panel), the recovery appears to
be good: at most scales, the recovery is accurate and has small
errors. At large scales the error bars increase in size because of
sample variance, and it appears that the recovered power spectrum lies
systematically below the input spectrum. This happens because at large
scales, we fit away some of the signal power during the foreground
fitting. If the points at large scales do not appear to jump around as
one would expect given the size of the error bars, this is because the
error bars here are dominated by sample variance, and so show our
uncertainty as to how representative this volume is of the whole
Universe. If, instead, we showed error bars showing only the
uncertainty on our determination of the power spectrum \emph{within
this volume}, they would be much smaller and would be visually
consistent with the scatter displayed by the points. The error bars
grow at small scales because the noise power becomes larger compared
to the signal power, limiting our sensitivity. We caution that, as
noted in Section~\ref{sec:sims}, the simulation we use represents a
rather optimistic scenario for low-redshift signal extraction, since
reionization occurs very late.

As we move to higher redshift (middle panel) the situation worsens
slightly, with the error bars increasing in size because of the higher
noise levels. More worryingly, the recovered power is lower than the
input power at all scales (though it becomes worse at large scales as
before) which seems to indicate that foreground subtraction may cause
significant bias in our estimate of the signal power even at
intermediate scales. The trend continues as we move to the highest
redshift slice (top panel). We do not plot the recovered power for a
range of scales between $k\approx 10^{-0.9}$ and $10^{-0.3}\ h\
\mathrm{Mpc}^{-1}$. This is because we infer an unphysical negative
power here. In the case of such points we plot a statistical upper
limit on the power. The bias from the fitting procedure leads to a
situation where these `upper limits' lie below the true power, or are
too small even to show up on the plot. These upper limits should, then,
be taken merely as an indication of the size of the fitting bias. The
larger noise at lower frequencies (higher redshifts) increases the size
of the error bars compared to the other panels. The combination of this
higher noise and the larger foreground power makes fitting the
foregrounds at high redshift more difficult, as we have seen in
previous work \citep{SKE_09,NONPAR_09}, leading to the observed bias.
\begin{figure}
  \begin{center}
    \leavevmode \psfig{file=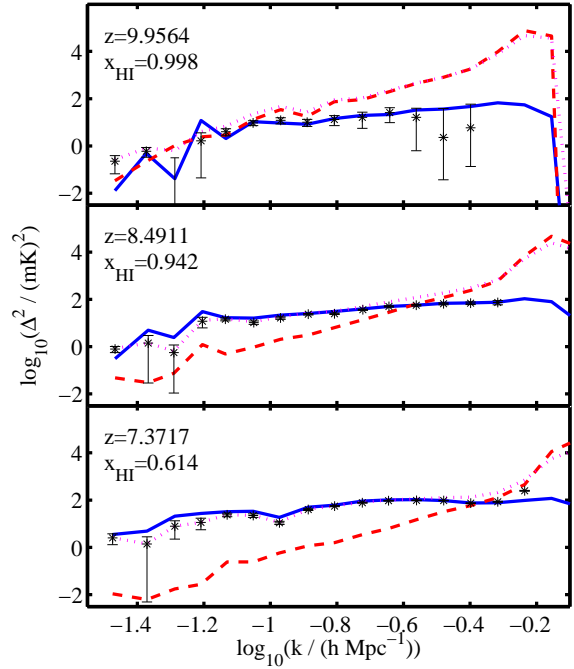,width=8cm}
    \caption{Power spectra of the CS, the noise, the residuals and the
    extracted signal for the case when the {\it uv} coverage is
    frequency-dependent, we have 300 hours of observation per
    frequency channel with a single station beam, and the foreground
    fitting is done using Wp smoothing in Fourier space. The redshift
    slices and the colour coding of the lines are the same as for
    Fig.~\ref{fig:threez300bin4}, but note we have changed the scale
    of the vertical axis to accommodate the upturn in noise power at
    small scales.}\label{fig:threezuvwp300}
  \end{center}
\end{figure}

The situation is very similar if the {\it uv} coverage is
frequency-dependent but we do our fitting using Wp smoothing in Fourier
space. This case can be seen in Fig.~\ref{fig:threezuvwp300}, which is
otherwise very similar to Fig.~\ref{fig:threez300bin4} except that we
have changed the vertical axis scale to accommodate the upturn in noise
power at high $k$ caused by the varying {\it uv} coverage. The higher
small-scale noise coming from the frequency-dependent {\it uv} coverage
damages the recovery of power at the smallest scales, but the fitting
using Wp smoothing in Fourier space allows us to recover the power on
intermediate and large scales even better than in
Fig.~\ref{fig:threez300bin4}. The reason that we fit even better than
in the supposedly more ideal case of Fig.~\ref{fig:threez300bin4} is
partly that the noise is normalized in image space to the expected
level for single-channel images (see Section~\ref{subsec:instresp}),
and so the increase in small-scale noise in the frequency-independent
case is compensated by a reduction in large-scale noise, improving
recovery there. It is also the case that our {\it uv} plane fitting is
more adaptive, applying less regularization at scales where the
foregrounds dominate and the noise is low. Unfortunately we do not yet
have a well-motivated method to choose the regularization parameter
$\lambda$ automatically rather than varying it by hand, but this result
suggests that finding a suitable method could yield even more
improvement in the quality of the fitting.

\begin{figure}
  \begin{center}
    \leavevmode \psfig{file=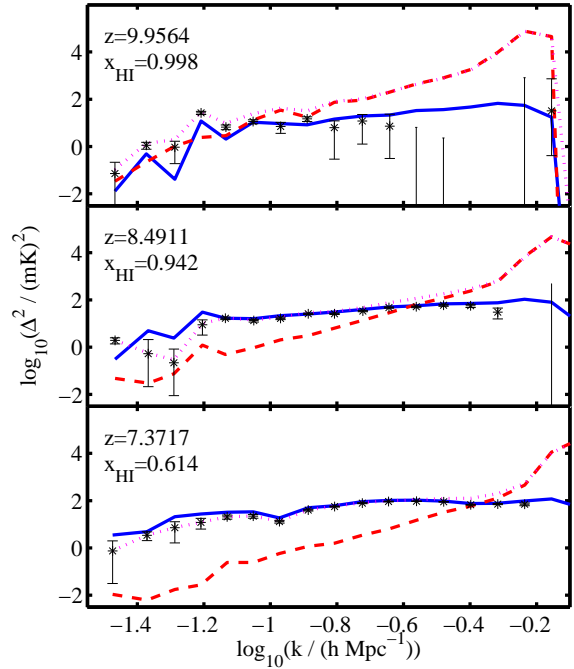,width=8cm}
    \caption{As for Fig.~\ref{fig:threezuvwp300}, except that the
    foregrounds are fit using a third-order polynomial rather than Wp
    smoothing.}\label{fig:threezuvpoly300}
  \end{center}
\end{figure}

If we use a third-order polynomial fit for the foregrounds rather than
using Wp smoothing, however, the result becomes worse, especially at
high redshift. This is illustrated in Fig.~\ref{fig:threezuvpoly300},
which is identical to Fig.~\ref{fig:threezuvwp300} apart from the fact
that polynomial fits are used.
While at low redshift the quality of recovery is visually
indistinguishable, at high redshift the Wp smoothing of
Fig.~\ref{fig:threezuvwp300} allows us to recover an estimate of the
power spectrum to higher $k$. The bias at low $k$ also seems to be
larger for polynomial fitting, which seems to produce overestimates of
the power of the CS at large scales. This may be due to the fact that
a polynomial is unable to match the large-scale spectral shape of the
foregrounds, allowing foreground power to leak into the
residuals. Unlike Wp smoothing, polynomial fitting does not allow us
to smoothly vary the level of regularization across the {\it uv} plane
(the only parameter we can tweak is the polynomial order, which is a
somewhat blunt instrument) and this may also contribute to the poorer
fit.

We conclude that even though varying {\it uv} coverage makes
foreground fitting more awkward, we can mitigate its effects without
having to discard a large proportion of our data if we choose our
fitting method carefully. At present our scheme for fitting the
foregrounds using Wp smoothing in Fourier space is quite slow,
however, so for the rest of the paper we revert to the case of
frequency-independent {\it uv} coverage, for which our image-space
fitting works quickly and reasonably
well. Fig.~\ref{fig:threezuvwp300} suggests that this should not
affect our comparisons of results using different lengths of observing
time or observational strategies. For actual LOFAR data, the fitting
of the foregrounds should still be much faster than other steps in the
reprocessing of the data, and so we are likely to use our most
accurate scheme (at present, Wp smoothing in Fourier space) even if it
is slow compared to other schemes.

\subsection{Different depths and strategies}

\begin{figure}
  \begin{center}
    \leavevmode \psfig{file=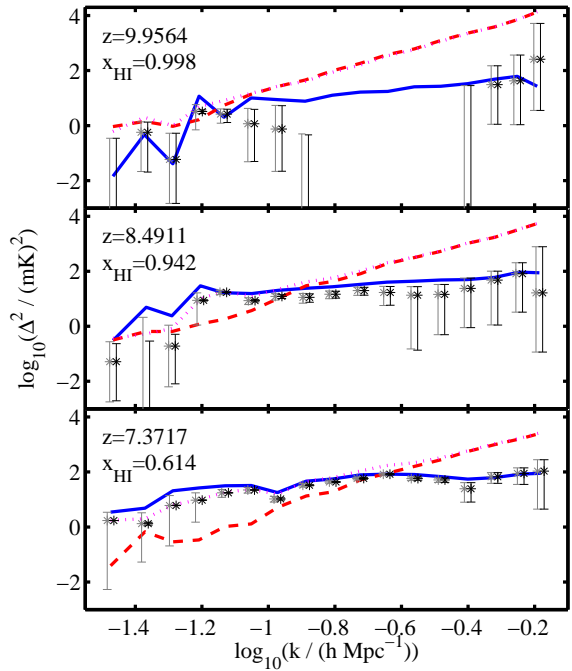,width=8cm}
    \caption{As for Fig.~\ref{fig:threez300bin4}, but using a noise
    level consistent with 180 hours of observation per frequency bin
    on a single window, using one station beam. We also plot two error
    bars for each point: the grey one on the left shows the error from
    both noise and sample variance as in our other figures, while the
    black one on the right shows the error only from
    noise.}\label{fig:threez180bin4}
  \end{center}
\end{figure}

Having compared the characteristics of different fitting methods, we
now move on to comparing the quality of extraction for different
assumptions about the amount of observing time, and for different
observational strategies. We start by showing the extraction for 180
hours of observing time per frequency bin, making a total of 360 hours
of observing time if two frequency ranges are required, in
Fig.~\ref{fig:threez180bin4}.
This makes it comparable to fig.~12 of \citet{BOW09}, who show a
simulated power spectrum for 360 hours of observation with the MWA
(though spanning a larger redshift range than a panel of our figure).
To make the comparison more illustrative, we show two error bars
for each point, the grey one on the left including both the noise error
and the sample variance, and the black one on the right including only
the noise error. For the MWA these would differ by less then ten
per cent and would be almost indistinguishable on this log-log scale
(J. Bowman, private communication). Visually, the errors for LOFAR
without sample variance appear smaller than those for the MWA at most
scales at the lower redshifts, as we may expect from the larger
collecting area. A computation including the sample variance, however,
tends to favour the MWA at small $k$ owing to its larger field of
view. Hence we explore the effect of observing multiple independent
windows below.

\begin{figure}
  \begin{center}
    \leavevmode \psfig{file=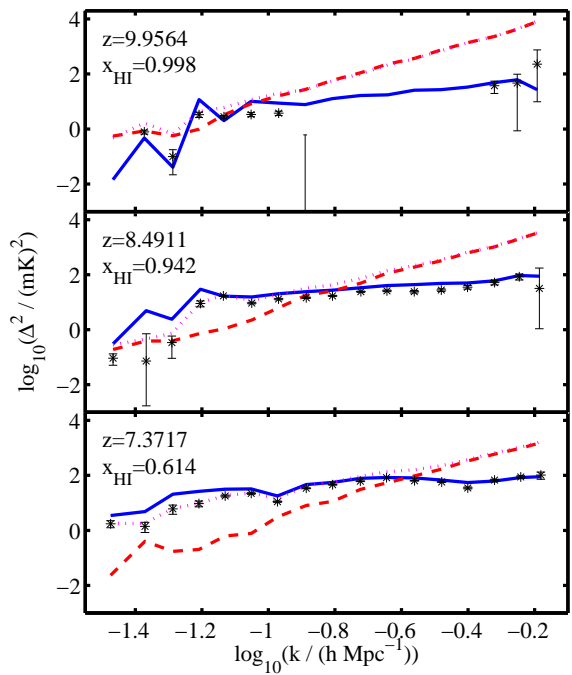,width=8cm}
    \caption{As for Fig.~\ref{fig:threez300bin4}, except we assume
    that six station beams are synthesized, rather than
    one.}\label{fig:threez6b300bin4}
  \end{center}
\end{figure}

The field of view can also be extended if, as planned, we are able to
synthesize multiple station beams simultaneously. Equivalently, if we
wish from the outset to observe a window larger than the $\sim
5\degr\times 5\degr$ of a single station beam, multiple beams can be
used to achieve observations of greater depth without using more
observing time. We show the effect of extending the field of view in
Fig.~\ref{fig:threez6b300bin4}, where we assume that we observe for
300 hours per frequency bin (as in Fig.~\ref{fig:threez300bin4}), but
using six station beams.
We model the effect of using six beams by reducing the errors due to
noise and to sample variance by a factor of $\sqrt{6}$. A realistic
primary beam model, and the incorporation of modes with smaller $k$,
would make the effect of multiple beams more complicated, but we
incorporate the effect in a way which is consistent with our
simplified beam. The most obvious effect of using multiple beams is at
large scales, since here the increase in the number of available modes
reduces the (large) sample variance errors as well as the noise
errors.  The noise errors at high $k$ are also reduced, however. Since
the smallest scales we probe may be comparable to the size of bubbles
in the H\textsc{I} distribution, this improvement may be important for
constraining physical models.

This figure also makes it clear what multiple beams do \emph{not}
do. Increasing the field of view in this way does not increase the
signal to noise along each line of sight, and so the foreground
fitting does not improve. The systematic offset at intermediate scales
in the middle redshift bin is still present, and we remain unable to
extract physically meaningful information at high redshifts at these
scales with our current methods. Our CS simulations are of limited
size, so we are unable to demonstrate how the larger field of view
enables us to recover the power spectrum at lower $k$. The bias we see
at the largest scales in our figures is unlikely to improve as we go
to yet larger scales, however, and so it may be difficult to exploit
the potential afforded by a larger field of view in practice.

We now directly examine the trade-off between spending observing time
to go deeper in a small area, and spending it to survey a larger
area. Considering first the situation at the lowest redshifts, we see
from Figs.~\ref{fig:threez180bin4} and~\ref{fig:threez6b300bin4} that
after 180 hours of observation per frequency channel, the fitting bias
has reached a level that reduces very little with deeper observation.
Moreover, with the six station beams of Fig.~\ref{fig:threez6b300bin4}
the errors at intermediate scales are rather small.  The main effect
of deeper observation is then to reduce the errors only at the very
smallest scales. It would clearly be more profitable to use extra
observing time to cover multiple windows, and reduce the large-scale
errors which are dominated by sample variance.

\begin{figure}
  \begin{center}
    \leavevmode \psfig{file=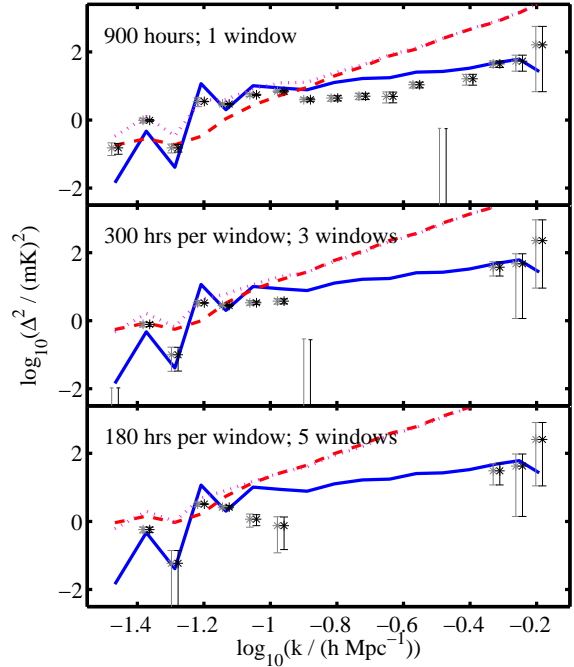,width=8cm}
    \caption{Power spectra of the original and extracted signal, the
    residuals and the noise, using the same line styles as
    Fig.~\ref{fig:threez300bin4}. Each panel assumes the same total
    observing time (900 hours) using six station beams, in an $8\
    \mathrm{MHz}$ slice centred at $z=9.96$ (with
    $\bar{x}_\mathrm{HI}=0.9976$), the same redshift as for the top
    panel of
    Figs.~\ref{fig:threez300bin4}--\ref{fig:threez6b300bin4}. The
    panels differ in the way in which the observing time is split
    between windows: in the top panel we devote all the observing time
    to a single window, and in the bottom panel we spread it equally
    between five different windows.  The middle panel shows an
    intermediate case. Each point has two error bars, the one on
    the right accounting only for noise, and the one on the left also
    including the effect of sample variance, as in
    Fig.~\ref{fig:threez180bin4}.}\label{fig:tradeoffi10}
  \end{center}
\end{figure}

At high redshift the trade-off between depth and number of windows is
more interesting, as we see in Fig.~\ref{fig:tradeoffi10}.
Here, all three panels show power spectra at the same redshift as the
top panel of our earlier figures ($z=9.9564$, with
$\bar{x}_\mathrm{HI}=0.9976$). Each point has two error bars, the
one on the right accounting only for noise, and the one on the left
also including the effect of sample variance, as in
Fig.~\ref{fig:threez180bin4}. The different panels distinguish between
different ways of allocating a fixed amount (900 hours) of observing
time per frequency band with six station beams. If we use this time to
observe five different windows (bottom panel), as seems to be
preferable at low redshift, the main effect is to reduce the size of
the statistical errors in a region of the power spectrum (low $k$)
where there is in any case a relatively large and uncertain systematic
correction to be made for the fitting bias. Meanwhile, the large amount
of noise per window degrades the fitting at intermediate scales. Taking
300 hours of observation per frequency band per window (middle panel)
reduces the bias somewhat, and enables recovery of reasonable quality
across a larger range of scales. Only with 900 hours of observation of
a single window (top panel), however, are we able to recover a
physically plausible estimate of the power across almost all the
accessible scales. Even at those scales at which the shallower
observations allowed some sort of estimate of the power, the increased
depth reduces the bias from the fitting, so that it becomes comparable
to the statistical error bars.

The tension between optimizing low- and high-redshift recovery is not
the only consideration in deciding how many windows to observe and for
how long. Using multiple windows will help to control the systematics
because we can then compare fields with different foregrounds and
different positions in the sky. If we wish to observe for a reasonable
fraction of the year, we are required to observe different windows
since some may be inaccessible or too low in the sky during some
periods. None the less, a hybrid strategy in which some windows
receive more time than others may be possible.

\begin{figure}
  \begin{center}
    \leavevmode \psfig{file=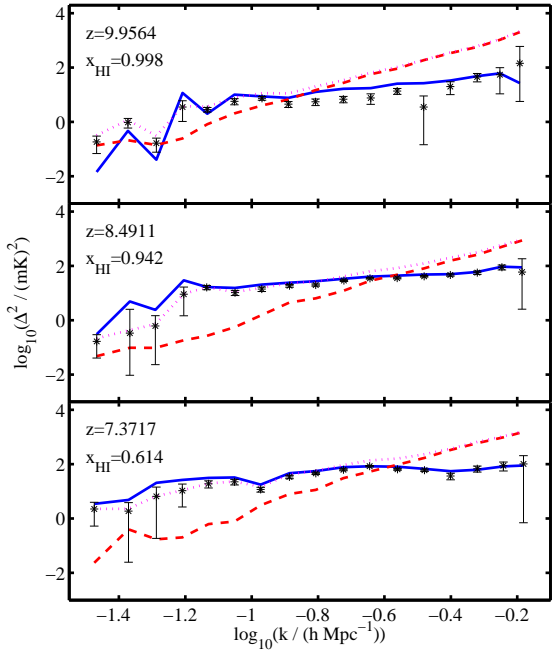,width=8cm}
    \caption{Power spectra at three different redshifts, using the
    same line styles as before. In this case, however, we assume that
    at frequencies above $160\ \mathrm{MHz}$ (corresponding to
    $z\approx 7.9$) we have used 300 hours of integration time, while
    below $160\ \mathrm{MHz}$ we have used 1200 hours of integration
    time, in each case using one station
    beam.}\label{fig:compositenoise}
  \end{center}
\end{figure}

Another possible strategy, since the higher redshift bins appear to
benefit more from longer integration times, is to spend longer
observing higher redshifts than lower redshifts. Since we already
split up the frequency range into different chunks which are not
observed simultaneously, this may be possible without excessive
difficulty. We note, however, that for other reasons (for example
improving the calibration), it may be desirable not to split the
frequency range into large contiguous chunks, but into two interleaved
combs. This would enforce a uniform integration time across the whole
frequency range. A further problem one may envisage is that the noise
rms would jump discontinuously across the gap between the two
frequency chunks. Unless the noise is well characterized, such a jump
could be confused with a change in the signal rms due to
reionization. It may also complicate the foreground fitting, and so we
test this in Fig.~\ref{fig:compositenoise}.
Here we have assumed that we have spent 1200 hours on the low
frequency chunk (below $160\ \mathrm{MHz}$), and only 300 hours on the
high frequency chunk. This does not appear to affect our fitting
adversely.  Even if we choose to plot the power spectrum in an $8\
\mathrm{MHz}$ slice which straddles the crossover between long and
short integration times, the extraction appears to be stable. If other
factors allow us to use such a strategy, then, it appears to be a
viable way to make the quality of our signal extraction more uniform
across the redshift range we probe.

\subsection{Source of the large-scale bias}\label{subsec:biassource}

\begin{figure}
  \begin{center}
    \leavevmode \psfig{file=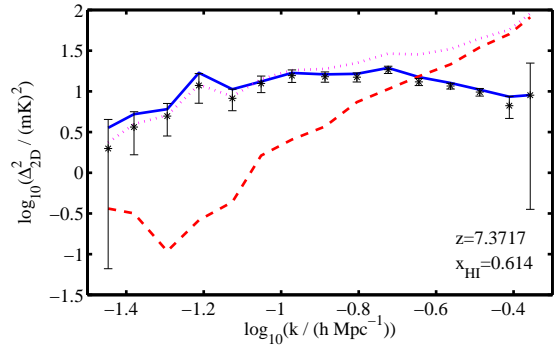,width=8cm}
    \caption{Two-dimensional power spectrum in the plane of the sky,
    for a slice $8\ \mathrm{MHz}$ thick centred at $z=7.3717$ and with
    $\bar{x}_\mathrm{HI}=0.6140$, for 900 hours of integration with a
    single station beam. The line styles for the original signal,
    noise, residuals and extracted spectrum are as for the previous
    figures.}\label{fig:p2d_i10_6b300}
  \end{center}
\end{figure}

\begin{figure}
  \begin{center}
    \leavevmode \psfig{file=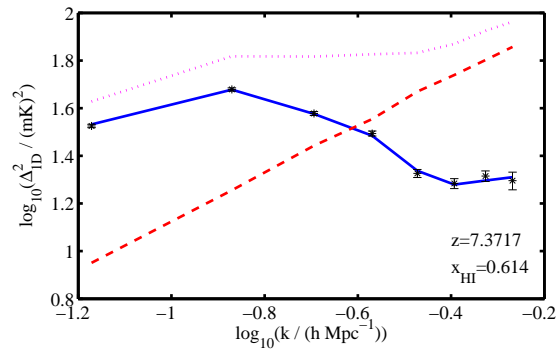,width=8cm}
    \caption{One-dimensional power spectrum along the line of sight,
    for a slice $8\ \mathrm{MHz}$ ($93.2\ h^{-1}\ \mathrm{Mpc}$) deep
    centred at $z=7.3717$ and with $\bar{x}_\mathrm{HI}=0.6140$, for
    900 hours of integration with a single station beam. The line
    styles for the original signal, noise, residuals and extracted
    spectrum are as for the previous
    figures.}\label{fig:p1d_i10_6b300}
  \end{center}
\end{figure}

Even when we achieve small statistical errors, as for the bottom panel
of Fig.~\ref{fig:threez6b300bin4}, a bias persists on large scales. We
look for the origin of this bias by plotting the power spectrum of
modes in the plane of the sky (the angular power spectrum) in
Fig.~\ref{fig:p2d_i10_6b300}, and the one-dimensional power spectrum
along the line of sight in Fig.~\ref{fig:p1d_i10_6b300}. For both of
these figures we consider a slice at low redshift (as for the bottom
panel Fig.~\ref{fig:threez6b300bin4}), and assume 900 hours of
observation per frequency chunk with one station beam.

The extracted two-dimensional power spectrum appears to behave
similarly to the three-dimensional power spectrum, albeit with
slightly larger error bars because we have fewer modes available. The
bias at large scales persists: we underestimate the power because we
fit away some of the signal and noise. The one-dimensional power
spectrum looks rather different.  It is quite accurately determined
because we average over so many lines of sight, and there is no
apparent bias in the extraction. The one-dimensional power spectrum
does not extend to such large scales as the two-dimensional power
spectrum because we restrict ourselves to quite a narrow frequency
slice (corresponding to a comoving depth of $93.2\ h^{-1}\
\mathrm{Mpc}$) to avoid evolution effects, but it does extend to
scales at which the two-dimensional power spectrum shows bias. We have
experimented with using slices which are twice as thick ($16\
\mathrm{MHz}$) and these still show no significant bias at the largest
scales. The one-dimensional power spectrum extends to smaller scales
than the two-dimensional one, since the spatial resolution is better
along the frequency direction for our $0.5\ \mathrm{MHz}$
channels. This resolution, and the lack of bias, may be useful if we
are able to invert the one-dimensional power spectrum to recover the
three-dimensional power spectrum \citep{KAI91,ZAR06}.

At first sight it seems somewhat puzzling that although we assume that
the foregrounds are smooth in the frequency direction -- we
effectively ignore very large-scale power along the line of sight --
the fitting bias manifests itself most clearly in the angular power
spectrum. Note, though, that if our estimate of the foregrounds along
a line of sight is offset by some constant, or by an amount that is
approximately constant within the narrow frequency range in which we
estimate the power spectrum (the fits are always computed across the
full frequency range to avoid edge effects), this does not change the
power spectrum of the residuals along the line of sight at all.  If
this offset is different between different lines of sight, though,
then this will be apparent in the angular power spectrum of the
residuals at each frequency. If the offsets at nearby points are
correlated, perhaps because the foregrounds within some region have a
similar shape and strength, then the angular power spectrum of the
residuals on small scales will hardly be affected. At scales larger
than the correlation length of the fitting errors then these offsets
could lead to the bias which we see.


In any case, Figs.~\ref{fig:p2d_i10_6b300} and~\ref{fig:p1d_i10_6b300}
suggest that we should consider the angular and line-of-sight power
spectra separately in an analysis of LOFAR data, though ultimately
neither will allow us to constrain models as tightly as a
three-dimensional power spectrum which includes a contribution from
all modes. The line-of-sight power spectrum appears to be less
vulnerable to bias and extends to higher $k$, while the angular power
spectrum extends to larger scales and may have greater power to
distinguish between models of reionization. The more sophisticated
version of this separation -- expanding the three-dimensional power
spectrum $P(k,\mu)$ in powers of $\mu$, the cosine of the angle
between a mode and the line-of-sight \citep{BAR05} -- is,
unfortunately, not likely to be useful for the noise levels expected
for LOFAR, though we have not yet made a quantitative investigation of
this possibility. \citet{PRI08} have checked this for an MWA-type
experiment, using an optimistic instrumental configuration, and find
that it does not have the required sensitivity. Rather, the separation
into powers of $\mu$ may have to wait for SKA or for a futuristic
lunar array.

\section{Summary and discussion}\label{sec:conc}

In this paper we have studied the extraction of the 21-cm EoR power
spectrum from simulated LOFAR data. The simulations allow us to
compute the statistical errors on the power spectrum due to thermal
noise and sample variance, and these are small enough to raise the
possibility of a significant detection of emission from the EoR using
only a modest amount of observing time. If we wish to estimate the
power spectrum accurately, however, this becomes more challenging once
we take into account the presence of fitting errors from the
subtraction of astrophysical foregrounds. These errors are correlated
(positively or negatively) with the signal and the noise in general,
and introduce a scale-dependent bias into our estimate of the power
spectrum. We anticipate that simulations such as the ones studied here
could be used to estimate and correct for the bias; this would induce
a further statistical error which can be straightforwardly computed by
using multiple realizations of a simulated observation. Making this
sort of correction will always be uncertain, though, so it is
desirable to minimize its size. We have looked at the extent to which
the size of the correction, as well as the size of the statistical
errors, can be reduced by observing for longer or using alternative
observational strategies.

Before that, though, we tested that extraction is still possible if we
do not make the assumption that the {\it uv} coverage is independent of
frequency. We find that this necessitates fitting the foregrounds in
the $(u,v,\nu)$ cube rather than the image cube, as noted by
\citet{LIU09b}. The Wp smoothing method, which we have used previously
to fit the foregrounds in the image cube, can be adapted to work in the
$(u,v,\nu)$ cube by fitting the real and imaginary parts independently
for each {\it uv} cell and by varying the regularization parameter,
$\lambda$, across the {\it uv} plane. This yields results comparable to
(in fact, even better than) those we obtain if we assume
frequency-independent {\it uv} coverage and then fit in the image cube.
We have also tried using a third-order polynomial to fit the
foregrounds in the $(u,v,\nu)$ cube: this yields results which are
acceptable, but not as good as those obtained using Wp smoothing. The
main drawback of Wp smoothing in this case is its speed, especially for
`lines of sight' near the centre of the {\it uv} plane where it is best
to choose a small value for $\lambda$ (implying little smoothing).
Because Wp smoothing in the image cube is faster, because the
polynomial fitting gives worse results than Wp smoothing in the
$(u,v,\nu)$ cube, and because Wp smoothing produces extraction of
similar quality in the image and $(u,v,\nu)$ cubes, we have
concentrated on results using frequency-independent {\it uv} coverage
to explore the different scenarios in this paper.

We have found that a year's observations (of, say, 600 hours, of which
perhaps 360 could be of a single window) should be sufficient to
detect cosmological 21-cm emission from towards the end of the EoR. We
caution, however, that the approximations employed in this paper
prevent us from treating these numbers as more than rough
estimates. If we wish to study the power spectrum at small or large
scales -- away from the `sweet spot' at intermediate $k$ -- it will be
important to be able to synthesize multiple station beams. This allows
us to reduce the statistical errors from sample variance and
noise. Unfortunately, however, there appears to be no substitute for
extending the integration time, especially to probe high redshifts and
very small scales. This is because only deep observations can improve
the quality of the foreground fitting, and hence reduce the systematic
offset between the true signal and the recovered signal.

Under the optimistic assumptions that we can synthesize six beams, and
that the useful frequency range can be covered using just two
frequency bands (the instantaneous frequency coverage is limited), 600
hours of observation of a single window should be enough to yield
quite precise and accurate power spectra up to $z\approx 9$, for $k$
between approximately 0.03 and $0.6\ h\ \mathrm{Mpc}^{-1}$. Pushing to
the very highest redshifts accessible with the frequency coverage of
LOFAR's high band antennas requires somewhat longer: perhaps 900 hours
per frequency band, which corresponds to 1800 hours of observation if
there are two frequency bands.

With observations of this depth, the limiting factor in the
statistical errors comes from sample variance on large scales, which
can only be reduced by observing a larger area of sky. This is one of
several reasons why the LOFAR EoR project plans to observe multiple --
perhaps five -- independent windows. We have already seen that
approximately 600 hours per window is required for the thermal noise
errors to be small and the bias to be under control for redshifts less
than about 9. For five windows, this corresponds to 3000 hours of
observation. Comparing the independent windows will also allow
important cross-checks, in particular that systematics are under
control.

To really push towards precise constraints on the power spectrum
towards the start of reionization, the 1800 hours per window that we
find yelds high quality extraction at $z>10$ corresponds to 9000 total
hours for five windows. This figure may be reduced if a hybrid
strategy, in which we integrate for a longer time in lower frequency
bands, turns out to be feasible. From the point of view of foreground
fitting and power spectrum extraction, ignoring constraints that may
be imposed by calibration etc., a hybrid strategy does indeed seem to
be feasible. Of course, we have considered this strategy only from the
point of view of the power spectrum.  If deeper observations at all
frequencies would allow us to push beyond the power spectrum, perhaps
into a regime where we can observe individual features in the
distribution of 21-cm emission towards the end of reionization with
reasonable signal to noise, then this would surely be valuable too.

Other hybrid strategies are also possible, for example ones in which
different windows are observed for different amounts of time. We have
not studied them here since they do not really impact the fitting and
extraction, which is independent for each window. None the less, they
may allow us to obtain high redshift constraints by observing one
window deeply, while simultaneously allowing us to beat down sample
variance errors on large scales at low redshifts by observing several
other windows at reduced depth.

In any case, our study suggests that as the amount of time spent
observing the EoR with LOFAR is increased, this allows us to make
qualitative improvements to the fitting, and to the range of scales and
redshifts we can probe accurately. Deeper integration does more than
simply allow us to shrink our statistical error bars.

This all depends, however, on the robustness of our fitting techniques,
and more generally on the level of control we are able to exercise over
systematic errors. The Wp smoothing method we have introduced
previously appears to work well when it comes to extracting the power
spectrum. This holds whether we apply it to an idealized case in which
the {\it uv} coverage of the instrument is constant with frequency, or
to a more realistic case in which it varies. We confirm a suspicion we
have expressed previously \citep{NONPAR_09} that the power spectrum may
be easier to extract than an apparently simpler statistic such as the
rms of the 21-cm signal: the fitting errors are scale-dependent, and a
power spectrum analysis allows us to pick out the scales where our
method works best without being swamped by small-scale noise. Splitting
the power spectrum into angular and line-of-sight components may help
us to test the robustness of our conclusions, and perhaps extend the
spatial dynamic range we can probe.

We have assumed here that the power spectrum of the noise is known to
reasonable accuracy, an assumption which will be examined in future
work. We will also study in a future paper how different strategies
alter our ability to constrain the parameters of reionization models.

Finally, we note that foreground fitting and power spectrum extraction
are late steps in the collection and analysis of LOFAR EoR data. They
depend on earlier and probably more difficult steps, such as
instrumental calibration (including polarization, which we have
neglected here), correcting for the ionosphere, and the excision of
RFI. The results of this paper only reassure us that the later stages
are unlikely to be the limiting ones.

\section*{Acknowledgments}

For the majority of the period during which this work was undertaken,
GH was supported by a grant from the Netherlands Organisation for
Scientific Research (NWO), and for the latter part of this period by
the LUNAR consortium (http://lunar.colorado.edu) which, headquartered
at the University of Colorado, is funded by the NASA Lunar Science
Institute (via Cooperative Agreement NNA09DB30A) to investigate
concepts for astrophysical observatories on the Moon. As LOFAR
members, the authors are partially funded by the European Union,
European Regional Development Fund, and by `Samenwerkingsverband
Noord-Nederland', EZ/KOMPAS. The dark matter simulation was performed
on Huygens, the Dutch national supercomputer. We thank the referee,
J. Bowman, for suggesting the cross-correlation technique of
Section~\ref{subsubsec:crosscorr}, and for improving the clarity of
the presentation.

\bibliography{allbib}

\end{document}